\documentclass[twocolumn,twoside,slac_two]{revtex4}
\usepackage{graphicx}
\usepackage{fancyhdr}
\begin{document}

\title{Recent Developments in Detector Technology}

%

\author{James E. Brau}
\affiliation{Center for High Energy Physics, University of Oregon, Eugene, OR 97403-1274,  USA}

\begin{abstract}
This review provides an overview of many recent advances
in detector technologies for particle physics experiments.
Challenges for new technologies include increasing spatial and temporal
sensitivity, speed, and radiation hardness while minimizing power and cost.
Applications are directed at several future collider experiments, including the
Large Hadron Collider luminosity upgrade (sLHC), the linear collider, and the super high luminosity B factory,
as well as neutrino and other fixed target experiments, and direct dark matter searches.  
Furthermore, particle physics has moved into space, with significant contributions
of detector technology, and new challenges for future efforts.
 
\end{abstract}

\maketitle

\thispagestyle{fancy}


\section{Introduction}

Advances in detector technology enable sharpened understanding and discovery
in particle physics.
As accelerator technology, and other innovations
such as detectors in space, advance the scientific frontier, it remains essential
that detector advances keep pace.  Detector advances also rejuvenate
the scientific programs of established accelerator facilities.

Scientific opportunities abound, but  pose great challenges for detectors.
The Large Hadron Collider luminosity upgrade (sLHC)\cite{slhc}, the International Linear Collider (ILC)\cite{ilc},
the Super B Factory\cite{superb}, neutrino experiments\cite{neutrino}, direct dark matter searches\cite{darkmatter},
and astroparticle physics experiments\cite{space} all will be limited
by the degree of success in detector R\&D.  Many have common challenges
and will benefit from parallel advances.

Significant effort is being devoted to developing detector technology.
This summary is necessarily incomplete due to 
time and space limitations, and the author apologies
for choices and omissions.

\subsection{Challenges}

The challenges to the experimenter are numerous.
Ever increasing precision in energy, momentum, space,
and time is needed.  Detectors must be fast, with minimal
occupancy.  Radiation hardness is demanded, 
as is background rejection.  Power and cooling constrain
detector design, and must be dealt with.  Cost
constraints impose practical limitations.
These and other considerations motivate efforts to
surpass all current performance limits.

Details on many
advances and developments in progress
are reported in contributions to
recent detector conferences, including the Dresden IEEE
Nuclear Science Symposium\cite{ieee}, TIPP09\cite{tipp09},
and the latest Pisa meeting\cite{pisa}.

\subsection{Enabling Advances}

Improvements
in detector technology often come from capitalizing on
industrial progress.  This happens now  at an
accelerated pace.
Ever finer feature size becomes possible, with detector elements of 10 $\mu$m or less
now feasible.  Likewise, faster,  low noise, low
power electronics is commonly available.  Integration advances
in microelectronics and mechanics enable novel, complicated
instrumentation.  New material developments create opportunities
for detector designs which are highly radiation resistant, robust,
thin, and durable.  Understanding
damage and annealing mechanisms increases hardness
to very high radiation dose levels.

\subsection{The Enterprise}

The particle physics detector community works at
several differing
scientific fronts, with
distinct challenges, requirements, and
solutions.  For colliders, 
detector applications  address tracking needs
for vertex and outer tracking systems, calorimeters,
and particle identification (including muon detectors).
Quite different challenges appear for direct
dark matter experiments.
In another thrust, neutrino detectors 
aim for the largest mass. Finally, the field has moved into
particle astrophysics experiments on Earth and in
space, creating yet additional issues.  
These diverse applications require many
different approaches.

Advances  build on the core technologies.
These include solid state tracking detectors, gaseous detectors,
crystals, 
cryogenic liquids, readout electronics, services
(including power, cooling, support, and materials), metrology,
and trigger and data acquisition.   Advances on all fronts are
needed, in parallel, to advance the scientific frontier.

\section{Overview of Experimental Programs}

A diverse portfolio of experimental programs drives detector advances.
These include the LHC experiments, the future linear
collider, heavy flavor experiments, neutrino experiments,  direct dark
matter detection experiments, and operations in space.

\subsection{LHC Detectors}

Successful construction and commissioning of the LHC detectors established 
critical lessons for the future.  The detectors are now operating, and planning is
underway for upgrades to deal with increasing LHC luminosity.  A luminosity of
$10^{35}$ cm$^{-2}$s$^{-1}$ is anticipated for the sLHC at 
the end of the decade.
Such a luminosity will require innovations, particularly for the inner trackers, where
a complete replacement will be required.  
Radiation damage limits are stretched, and rates exceed the
limits for some candidate technologies, requiring increased granularity.  In addition to
upgrades for the inner tracking, other systems will also need to be improved,
especially the electronics.

\subsection{Linear Collider Detectors}

The linear collider detectors will require exceptional precision,
with time stamping to suppress backgrounds.  The ILC is designed
for a bunch train of about about 3000 bunches over 1 msec.  
The goal for vertex detector precision is
$<$4 $\mu$m, based on  $\sim$20 $\mu$m pixels
in order to optimize heavy flavor tagging.
Momentum resolution must be
 $\sigma$(1/p) $\sim$ few $\times 10^{-5}$
to reconstruct the Higgs as a missing mass in the Higgstrahlung channel.
Physics requirements (W/Z separation in the dijet channel) set the jet resolution requirement of
$\sigma$(E$_{jet}$)/E$_{jet} \sim$ 3-4\%  for E$_{jet} >$ 100 GeV.

\subsection{Heavy Flavor Experiments}

LHCb will soon commence
operation.  It will be challenged by high radiation
levels, a particular problem for the vertex locator (VELO),
which is expected to eventually need replacement.

Efforts toward a Super B Factory aim at reduced
multiple scattering in the tracker with thinner detectors,
rad-hard endcap crystals, and advanced end-cap particle
identification techniques.

NA62  aims to measure the rare decay
$K^+ \rightarrow \pi^+ \nu \overline{\nu}$,  having
developed a state-of-the-art giga-tracker and a
RICH designed to suppress the $K^+ \rightarrow \mu^+ \nu$
background by three orders of magnitude (after a kinematic
suppression of five orders of magnitude and another five
orders of magnitude suppression from penetration measurements.)

MEG employs a liquid xenon calorimeter in the search for $\mu \rightarrow$  e $\gamma$.
It demands high purity and good understanding of the response and calibration.

\subsection{Neutrinos}

Neutrino experiments are benefiting from recent technological
advances.  The development of multi-pixel photon counters (MPPCs, or
silicon photomultipliers - SiPMs) is an important one.  T2K employs about 60,000
devices to read out the scintillator bars in the near detector.
NOvA is constructing a massive (16 kton) extruded CVC contained
liquid scintillator detector in search of $\nu_\mu \rightarrow \nu_e$.

Looking toward the future, megaton-scale detectors are envisioned.
For the full scale water Cherenkov detector, innovations in
photomultiplier tubes to manage costs are needed.
A somewhat less massive detector with tracking liquid argon may achieve
the physics goals.  

\subsection{Direct Dark Matter Detection}

Direct dark matter detection experiments search for very
weakly interacting particles, which requires extremely sensitive detector elements, 
with a threshold of just a few keV.  Background suppression
is essential, driving the experiments deep underground, with passive 
shields, and detectors built with low intrinsic radioactivity, and capable
of gamma background discriminatation.  Experimenters look for signatures
based on ionization, scintillation, phonons, or a combination.

\section{Technologies}

Advances in detectors span a wide spectrum of technologies.
There are notable recent advances in silicon and diamond tracking 
detectors, gaseous tracking detectors, electromagnetic and hadronic calorimeters,
detectors for particle identification, progress toward megaton neutrino detectors, 
and dark matter detectors.

\subsection{Solid State Tracking Detectors}

\begin{figure}[h]
\centering
   \includegraphics[width=80mm]{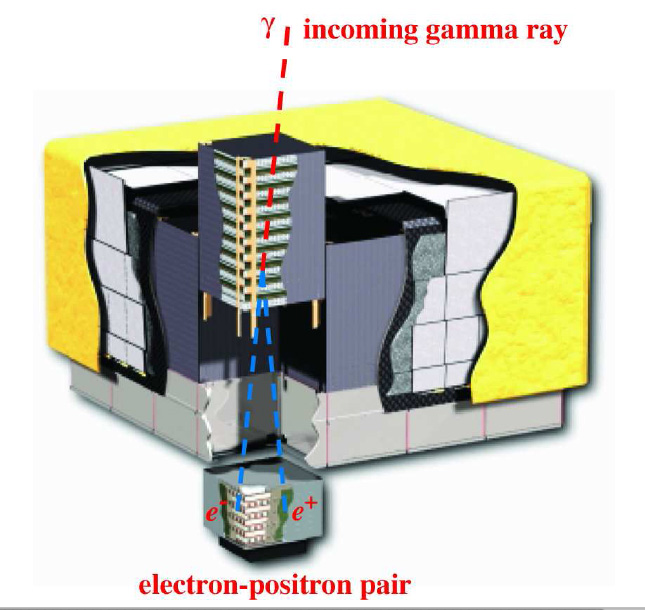}
 \caption{Schematic diagram of the Fermi Large Area Telescope\cite{fermi}.
The converter-tracker has 16 planes of high-Z material interleaved with tracking 
planes of single-sided silicon strip detectors.  The overall dimensions of
the telescope are 1.8 m $\times$ 1.8 m $\times$ 0.72 m. 
 }
 \label{fermi}
\end{figure}

The construction and commissioning experience of the LHC detectors\cite{atlas,cms} and the Fermi
Gamma-ray Space Telescope\cite{fermi}(Figure \ref{fermi}) provided many lessons for large scale silicon
systems.  The next challenges will include the increased rate and radiation levels of
the sLHC, the increase precision demanded by the ILC and the B Factories,
and a variety of specialized applications, such as the NA62 Gigatracker.

\begin{figure}[h]
 \begin{center}
   \includegraphics[width=80mm]{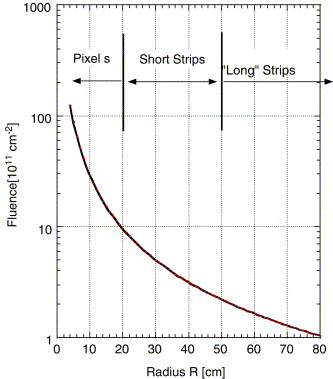}
 \end{center}
 \caption{Expected radial fluence as a function of radius for an sLHC detector after an integrated luminosity of 2500 fb$^{−1}$,
with radial extent of proposed tracker regions. \cite{slhc_track}
 }
 \label{trkrad}
\end{figure}

\begin{figure}[h]
\centering
   \includegraphics[width=80mm]{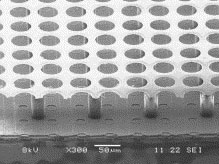}
 \caption{The Integrated Grid (InGrid) on top of a 20 $\mu$m SiProt layer, on top of a TimePix chip\cite{gossip}.
 }
 \label{gossip}
\end{figure}

\begin{figure}[h]
\centering
   \includegraphics[width=80mm]{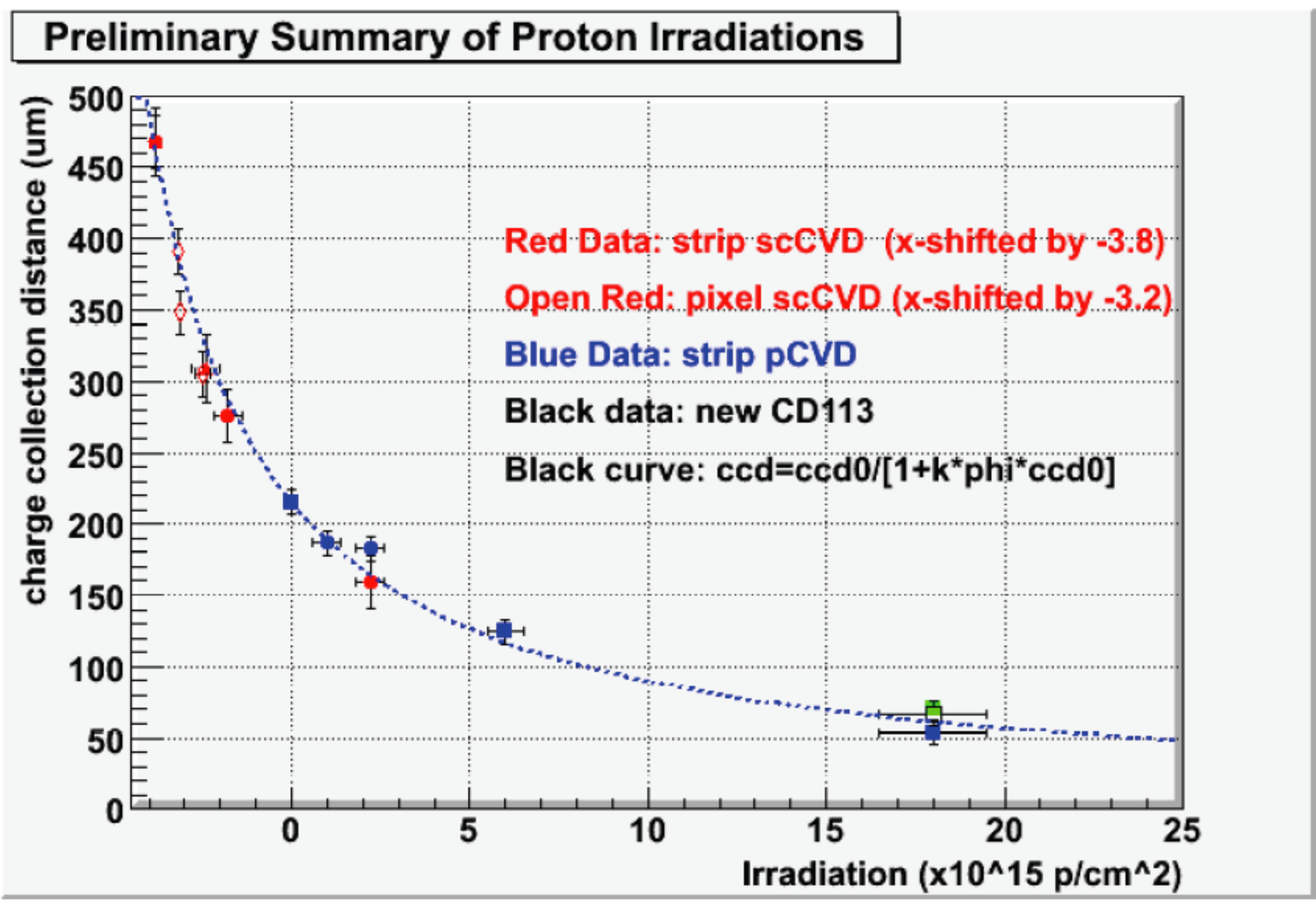}
 \caption{Proton irradiation results for pCVD (blue
points) and scCVD (red points) material at 1V/m
and 2V/m (green square). See reference \cite{diamond}
for interpretation.
 }
 \label{diamond}
\end{figure}

The event rate at the sLHC will be about 300-400 events per crossing, 
with about 10,000 tracks within $|\eta| \le$ 3.2.
The intense radiation levels\cite{slhc_track} posed by the sLHC for tracking are
 $10^{16}$ /cm$^2$  ($\sim$400 MRad) at a radius of 5 cm from the beamline,
 $10^{15}$ /cm$^2$ ($\sim$40 MRad) at a radius of 20 cm, and
 2 $\times$ $10^{14}$ /cm$^2$  ($\sim$10 MR) at a radius of 50 cm
(See Figure \ref{trkrad}).
These levels dictate the technologies under consideration\cite{slhc_track}.
Beyond 60 cm conventional silicon strips are suitable, but within 60 cm
short strip detectors must be used.  Within 20 cm pixel technologies
are required, while within $\sim$10 cm,
even the standard pixel
technology is unacceptable, and new technologies are being developed.
ATLAS considers several candidate technologies for this innermost region,
including
planar technology, 3D-silicon, diamond\cite{diamond}, and GOSSIP\cite{gossip}, a gaseous pixel detector (see Figure \ref{gossip}).
Radiation damage studies of diamond are presented in Figure \ref{diamond}.

Development of silicon sensors for linear collider vertex detectors
confront a different set of requirements.  Radiation levels 
will be much lower, but the physics and machine demands
call for
\begin{itemize}
\item excellent spacepoint precision ($<$ 4 microns),
\item superb impact parameter resolution (5$\mu$m $\oplus$ 10$\mu$m/(p sin$^{3/2}\theta$)),
\item transparency ($\sim$0.1\% X$_0$ per layer),
\item track reconstruction (find tracks in the vertex detector alone),
\item sensitivity to minimal bunch crossings ($<$150 crossings = 45 $\mu$sec for ILC),
\item EMI immunity, and
\item limited power($<$100 Watts for passive cooling to minimize material).
\end{itemize}

Several efforts are addressing these challenging requirements.
Concepts under development include charge-coupled devices (CCDs),
which were successfully applied in the 307 Megapixel SLD vertex 
detector\cite{sld}, as well as several other approaches.
CCDs as implemented by SLD are too slow for the ILC.
The efforts to advance the technology include\cite{ilcvxd}:
\begin{itemize}
\item column parallel readout CCDs,
\item FPCCDs (fine pixel CCDs),
\item monolithic active pixels based on CMOS 
(MIMOSA, MAPs, FAPs, Chronopixels, 3D-SOI),
\item DEPFET (DEpleted P-channel Field Effect Transistor ),
\item SoI (silicon on insulator),
\item ISIS (Image Sensor with In-Situ Storage), and
\item HAPS (hybrid active pixel sensors).
\end{itemize}


Silicon tracking is also under development for the linear collider
by the SiD design group\cite{sid}.
The superb resolution of silicon allows a small tracking volume
to achieve the tracking goals of the ILC, better than $\sigma_p$/p $=$ 1\% 
at 100 GeV/c.
Silicon is fast, and provides robustness to backgrounds
and occasional beam disruptions anticipated at the linear collider.
The performance goal requires very low mass support so as
to minimize the material within the tracking volume.  
The SiD design of modular low mass sensor tiles of carbon
fiber cylinders achieves 0.6\% of a radiation length per
layer, with passive cooling. The anticipated material budget
of the inner tracking system (vertex and tracker) is less than
10\% of a radiation length for $\theta >  50 ^\circ$.

A silicon envelope for a TPC tracker is being studied by the
SiLC Collaboration\cite{silc}.  This system is envisioned to
provide improved tracking performance, tracking hermiticity,
monitoring and alignment, and robustness.


The Gigatracker\cite{na62} under development for NA62 employs
three silicon pixel sensors to provide precise direction and timing
information.  A gigaHertz rate with more than a megaHertz/mm$^2$ 
at the maximum must be handled in vacuum.  Two readout
options are being developed, a constant fraction discriminator (CFD)
with complex pixel circuitry, and a time over threshold (TOT) with
simple, low power pixel circuitry. Acceptable prototypes of the analog circuits
for both options have been produced in 0.13 $\mu$m CMOS.


Diamond has certain properties that give it a potential advantages over silicon as a sensor
material.  These include a larger bandgap and the consequent low
leakage current, the  strong atomic bonds which make for high radiation hardness,
and the smaller dielectric
constant.  
Chemical vapor deposition diamond is available as
single-crystal (scCVD) or polycrystalline (pCVD) diamond.
pCVD wafers can be grown to
12 cm diameter and 2 cm thick.  
scCVD diamonds with an area
of a few cm$^2$ and about 1 mm thickness are available.
Over the past several years, experience with diamond
as radiation monitors have been achieved, and it now appears
as a candidate for the LHC inner tracking (mentioned above)\cite{diamond}.

\subsection{Gaseous Tracking Detectors}

The ALICE time projection chamber (TPC) is the largest
TPC ever built.  It has an outer radius of 2466 mm, and a
drift length of 2 x 2500 mm drift\cite{matyja}.

\begin{figure}[h]
 \begin{center}
   \includegraphics[width=80mm]{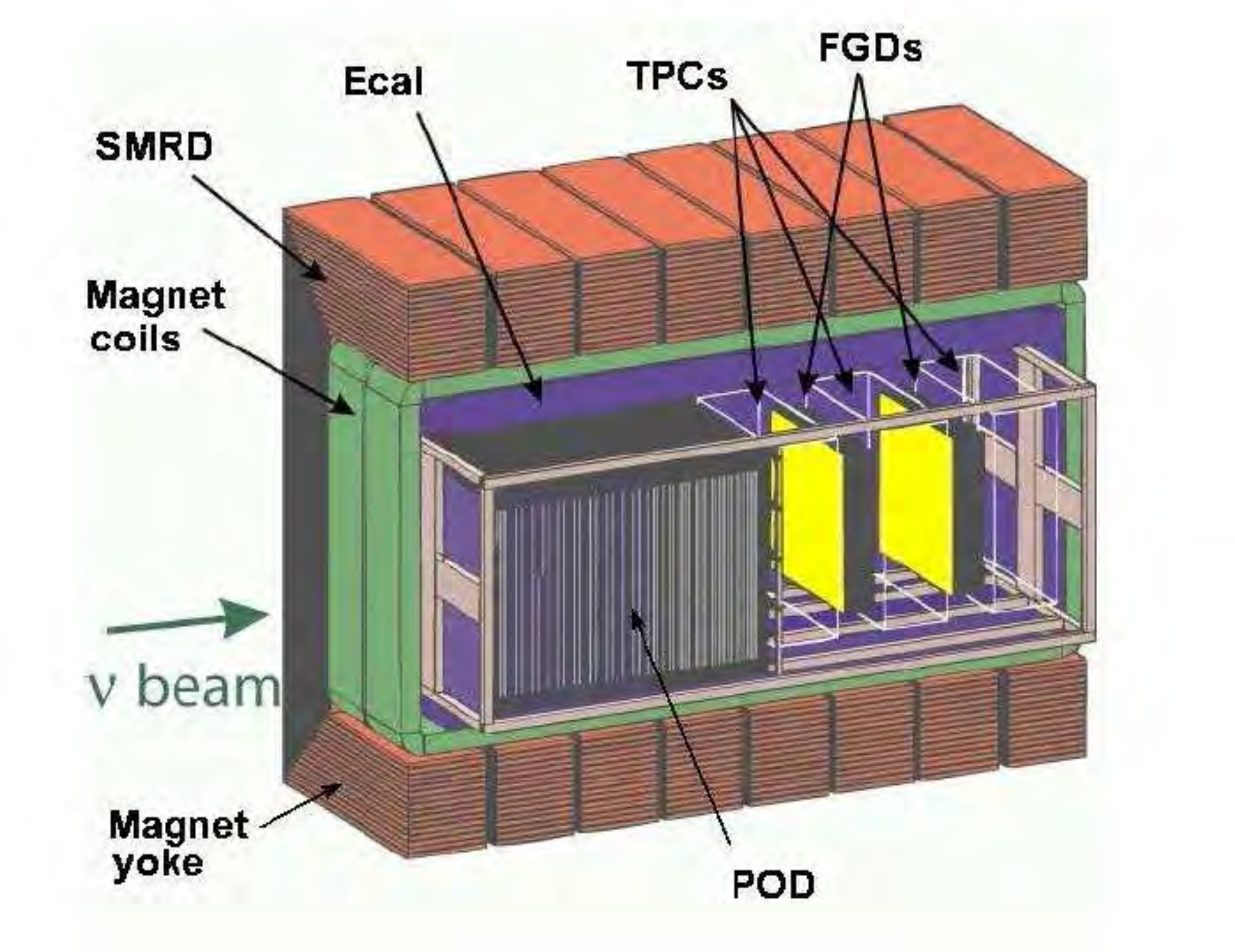}
 \end{center}
 \caption{Cutaway view of the T2K off-axis near detector\cite{giganti}.
 }
 \label{t2k}
\end{figure}

Recent work has been aimed at 
developing micropattern gas detectors (MPGDs)\cite{mpgd} as
replacements for the conventional
gaseous TPC readout.
MPGDs are immune to E $\times$ B effects that plague conventional detectors. 
This work includes GEMs, MicroMeGas, and the Timepix (CMOS)/Ingrid approach.
To date, the T2K off-axis near detector is the largest TPC equipped with MPGDs.
The three
T2K TPCs (Figure \ref{t2k}) consist of 72 Micromegas modules, 12 modules per readout
plane, with  9 m$^2$ active area comprising 120k electronics channels\cite{giganti}.
The magnetic field is 0.2 Tesla.

The LCTPC Collaboration is studying a large TPC 
(720 mm inner diameter field cage) with MPGDs
in beam\cite{dehmelt} as a prototype for an ILC detector\cite{ild}.
Both double GEMS and Micromegas have
been tested within a 1 Tesla field. A triple GEM structure with Timepix
readout has also been tested\cite{kaminski}.

\subsection{Electromagnetic Calorimetry}

Notable recent work on electromagnetic calorimetry has included
silicon-tungsten, scintillator strips, crystals, and liquid xenon for MEG.


A fine grained electromagnetic calorimeter serves several  goals  toward  optimized  
physics  performance of the ILC experiments.  It is  critical to the success  of the particle  
flow  technique  for  hadron  jet  calorimetry,  but also important  for other  physics, 
such as the tau decay reconstruction.
A natural technology choice is the silicon-tungsten (SiW) sampling calorimeter.   
Good success was achieved using SiW for luminosity monitors at SLD,  DELPHI,  
OPAL,  and ALEPH \cite{Berridge:1992dw, Strom:1993jv, Bederede:1995pc,
Abbiendi:1999zx, Almehed:1991tk}.   SiW sampling calorimetry specifically optimized for the 
ILC experiments is under development by two groups, the SiD design group, and the 
CALICE Collaboration.

CALICE has been conducting a series of beam tests over the past few years\cite{carloganu}.
Beginning at DESY in 2006, the tests continued at CERN (2006-7), and at Fermilab
in 2008-9.  The data from 6 GeV to 25 Gev shows a resolution of

\begin{eqnarray*}
   {\sigma_E \over E} = {(16.53 \pm 0.14(stat) \pm 0.4(syst)) \% \over \sqrt{E(GeV)}}   \\
   {} { \oplus (1.07 \pm 0.14(stat) \pm 0.1(syst))\%}
\end{eqnarray*}

\noindent The collaboration is also building a technological prototype, which will
integrate the mechanical, thermal, and electric constraints for an ILD electromagnetic
calorimeter module,
include embedded readout electronics, and allow for a realistic study of the
construction processes.

The SiD team (a SLAC/Oregon/BNL/Davis/ Annecy Collaboration) is developing 
a very dense, fine grained silicon tungsten calorimeter for the ILC, with a pad size 
of 13 mm$^2$ to match the Moliere radius ($\sim$ R$_M$/4)\cite{sidsw}. Each six inch wafer of 1024 pads 
is read out by one integrated chip (KPiX).  Tests have demonstrated that less than 
1\% crosstalk is expected.  A KPiX design\cite{kpix}
 has been fabricated and is under test.  
In order to be sensitive to single minimum ionizing particles, the signal-to-noise 
must exceed seven, making the required noise level less than 2000 electrons.  
A dynamically switchable feedback capacitor scheme achieves the required dynamic range 
of  0.1-2500 MIPs.  The mechanical design calls for passive cooling, with heat 
conducted through the tungsten to the edge of each module.  
The collaboration is working toward a prototype tower of the pad
structure. A  MAPS version is also under development in collaboration with RAL\cite{Ballin:2009yv}.

The CALICE Collaboration has tested a scintillator strip electromagnetic calorimeter
read out by multi-pixel photon counters in a beam at DESY and Fermilab.  It is designed
for the particle flow technique at the ILC, where
3 - 5 mm strips for high granularity are needed.

Crystals have a long history in electromagnetic calorimetry\cite{crystal_history},
and new HEP calorimeter applications
are now under development. Applications include
 PWO for PANDA at GSI,
LYSO  (Cerium doped Lutetium Yttrium Orthosilicate, 
Lu$_{2(1-x)}$Y$_{2x}$SiO$_{5}$:Ce)
for a Super B Factory, Mu2e and the CMS endcap upgrade,
            and PbF$_2$, BGO, or PWO for a homogeneous hadron calorimeter.
The SuperB/Mu2e/CMS endcap application demands the highest possible
radiation hardness.  LYSO
is currently favored for its radiation hardness, large light yield,
and low noise.  The homogeneous hadron calorimeter (HHCAL) application is an interesting
recently advanced crystal concept\cite{hhcal} describe below.

The MEG experiment at the PSI employs an
800 liter liquid xenon calorimeter, with  846 PMTs.
It now starts a new run with improved performance.
Superb resolutions in energy, position, and time are possible\cite{MEG}.

\subsection{Hadronic Calorimetry}

\begin{figure}[h]
 \begin{center}
   \includegraphics[width=80mm]{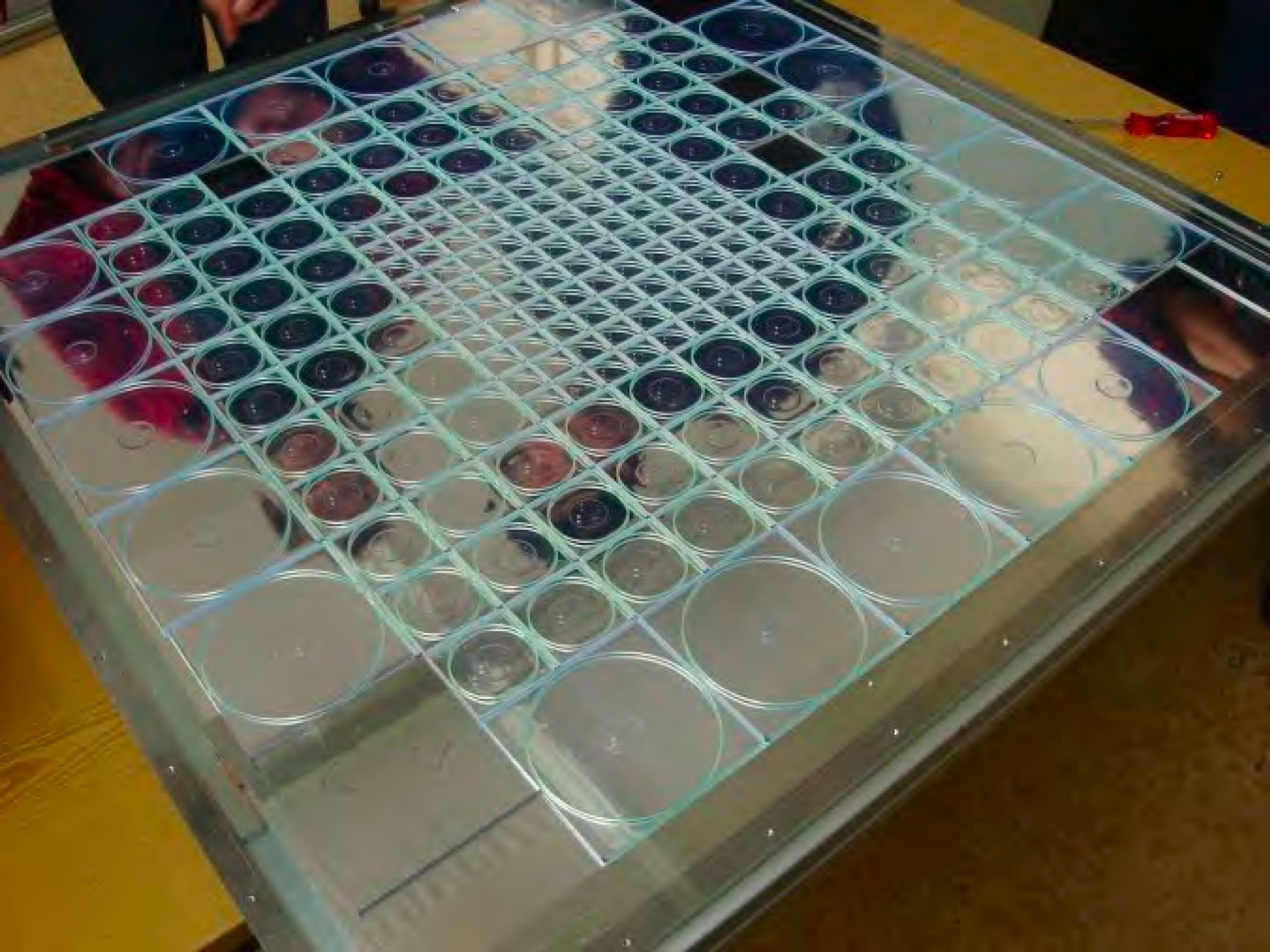}
 \end{center}
 \caption{CALICE scintillator hadron calorimeter layer. \cite{calicehad}.
The pattern is built from 216 tiles.
}
 \label{calicehad}
\end{figure}

Recent interest in particle flow calorimetry for the ILC drives advances on several 
technologies.  The fundamental idea of particle flow calorimetry is to capitalize
on the excellent tracker momentum measurement
of the charged component of a
jet, and to
measure only neutral components of the jet in the calorimeter.
These separately measured components can be combined for
an optimal jet energy measurement.  This technique requires
excellent separation of the calorimeter depositions from
neutral and charged particles
with a highly granular calorimeter.
The hadron calorimeter detector technologies of interest in particle flow calorimetry include analog scintillator calorimeters
with silicon photomultiplier (SiPM) readout, and digital gas calorimeters, based on RPCs,
GEMs, or MicroMeGas\cite{sid, ild}.
The CALICE Collaboration has constructed and tested a scintillator calorimeter consisting
of 38 steel layers (2 cm thick) of about 1 m$^2$ for 4.5 interaction lengths, with 7608 scintillator tiles
read out with SiPMs. Each gap in the steel is instrumented with the highly 
segmented scintillator; the first 30 layers are assembled from tiles varying in 
area from 3 $\times$3 cm$^2$ at the center to
12 $\times$ 12 cm$^2$ (Figure \ref{calicehad}).
The calorimeter has been tested in beams at DESY, CERN,
and Fermilab\cite{calicehad}. The tests included running with a silicon-tungsten
electromagnetic calorimeter, and a tail catcher, muon tagger (TCMT) backup,
and featured common readout electronics.  The CALICE
Collaboration draws the following conclusions from these tests:
1.) the SiPM technology has proven to be robust and stable, 2.) the calibration is well under control,
3.) the performance is as expected and understood, and 4.) the results strongly support 
the predicted particle flow performance.

A small glass RPC calorimeter 
has been tested at Fermilab\cite{rpc}, 
and a large calorimeter designed for particle flow is 
under construction. The tested RPC module was built of
20 x 20 cm$^2$ RPCs with
	1 $\times$ 1 cm$^2$ readout pads.
The 10 chambers consisted of 2560 readout channels.
The beam tests involved 120 GeV protons, and 
 1-16 GeV $\pi^+$ and e$^+$.
A one m$^3$ prototype is under construction,
with beam tests and analysis
planned in 2010-11.

\begin{figure}[h]
 \begin{center}
   \includegraphics[width=80mm]{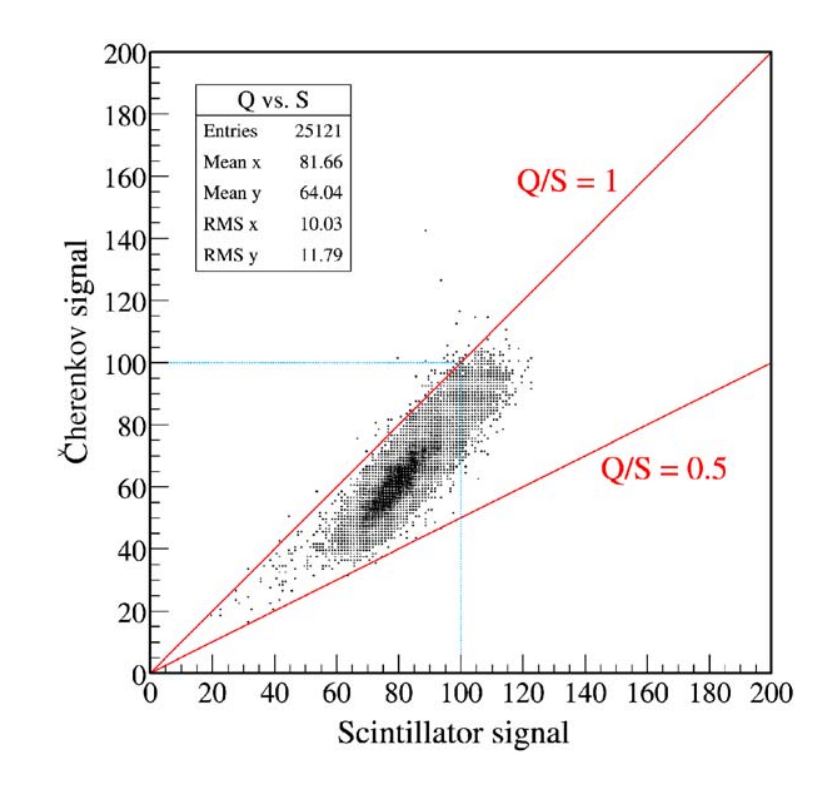}
 \end{center}
 \caption{Cherenkov signals versus scintillator signals for
100 GeV $\pi^-$ in the DREAM Calorimeter. \cite{dream}.
 }
 \label{dream}
\end{figure}

Since the fluctuations in hadronic showers are dominated by
nuclear binding energy losses and $\pi^0$ energy
fluctuations, the resolution of a hadron shower can be
improved by separately measuring the energy deposited
by electromagnetic showers resulting from the $\pi^0$s
from the energy of non-relativistic particles\cite{dual}.
The DREAM Collaboration has realized a measurement
of this with a copper fiber calorimeter, where quartz and scintillating
fibers separately measure these two components\cite{dream}.
While the constructed calorimeter is leakage limited for high 
energy showers, the collaboration is able to demonstrate improved
resolution when the two components are combined.
Figure \ref{dream} shows the response in the DREAM calorimeter
to 100 GeV$\pi^-$.  The Cherenkov and scintillator signals are shown
to respond differently to the electromagnetic content of the showers.

The homogeneous hadron calorimeter (HHCAL)\cite{hhcal} would operate in the dual readout mode,
separating the signals from relativistic particles and non-relativistic particles
by detecting Cherenkov and scintillation light separately.  Large volume is
required, so cost effectiveness is essential.  The crystal material must be UV transparent.
Three candidates have been evaluated (PbF$_2$, BGO, PWO), and
initial investigation favors scintillating PbF$_2$\cite{zhu}.

The sLHC's  harsh radiation environment will
pose an increased challenge for the forward
calorimeters. For example, the liquid argon in the ATLAS forward calorimeter
will heat and may boil, and space charge will build up between electrodes.
These effects could require new forward calorimeters.  
ATLAS is confronting this challenge with the 
development of two possible solutions.  A new forward calorimeter,
with smaller gaps and increased cooling would replace the existing
calorimeter, to handle the radiation.  Such a calorimeter is under development\cite{fcal}.
This solution requires opening the
cryostat.  An alternative solution would be to insert a warm calorimeter
in front of the current calorimeter.

\subsection{Particle Identification}

Particle identification has been an important element in
recent experiments (BaBar, Belle, LHC-b), and
continues to be a focus in detector development for 
future heavy flavor experiments (Belle II\cite{dolezal}, INFN Super B, NA62).  The key technologies are
radiators and photodetectors.  Polishing techniques are advancing
for quartz radiators, and silica aerogel  developments include
multi-index tiling. The photodector technology is advancing,
including the hybrid photodectors, microchannel plates with phototubes,
and multi-pixel photon counters (MPPCs or SiPMs).  The MPPCs
are single photon sensitive devices, built from avalanche photodiodes (APD)
arrays on a common silicon substrate.  They are extremely compact, B field
immune, provide very good timing, and offer gain and quantum efficiency
competitive with a PMT.  There are many developments and investigations
pushing this technology.  	Notable are the MPPCs for the fiber readout 
of the T2K near detectors, fabricated by Hamamatsu,
which contain 667 pixels 
(50$\times$ 50 $\mu$m$^2$)  on 1.3 mm $\times$ 1.3 mm arrays.  More than 60,000 arrays will
be used to read out all detectors except the TPCs\cite{t2kmppc}.

\subsection{Megaton Detectors for Neutrinos}

The strength of the water Cherenkov technique is well established by the
success of Super Kamiokande, and its predecessors.  The future goal,
aimed at super sensitive neutrino mixing and proton decay experiments,
is the megaton detector.  The challenges are costs, improved quantum
efficiency for the PMTs, readout electronics, new photosensors, and
hardening against unexpected accidents.  

\begin{figure}[h]
 \begin{center}
   \includegraphics[width=80mm,keepaspectratio=true]{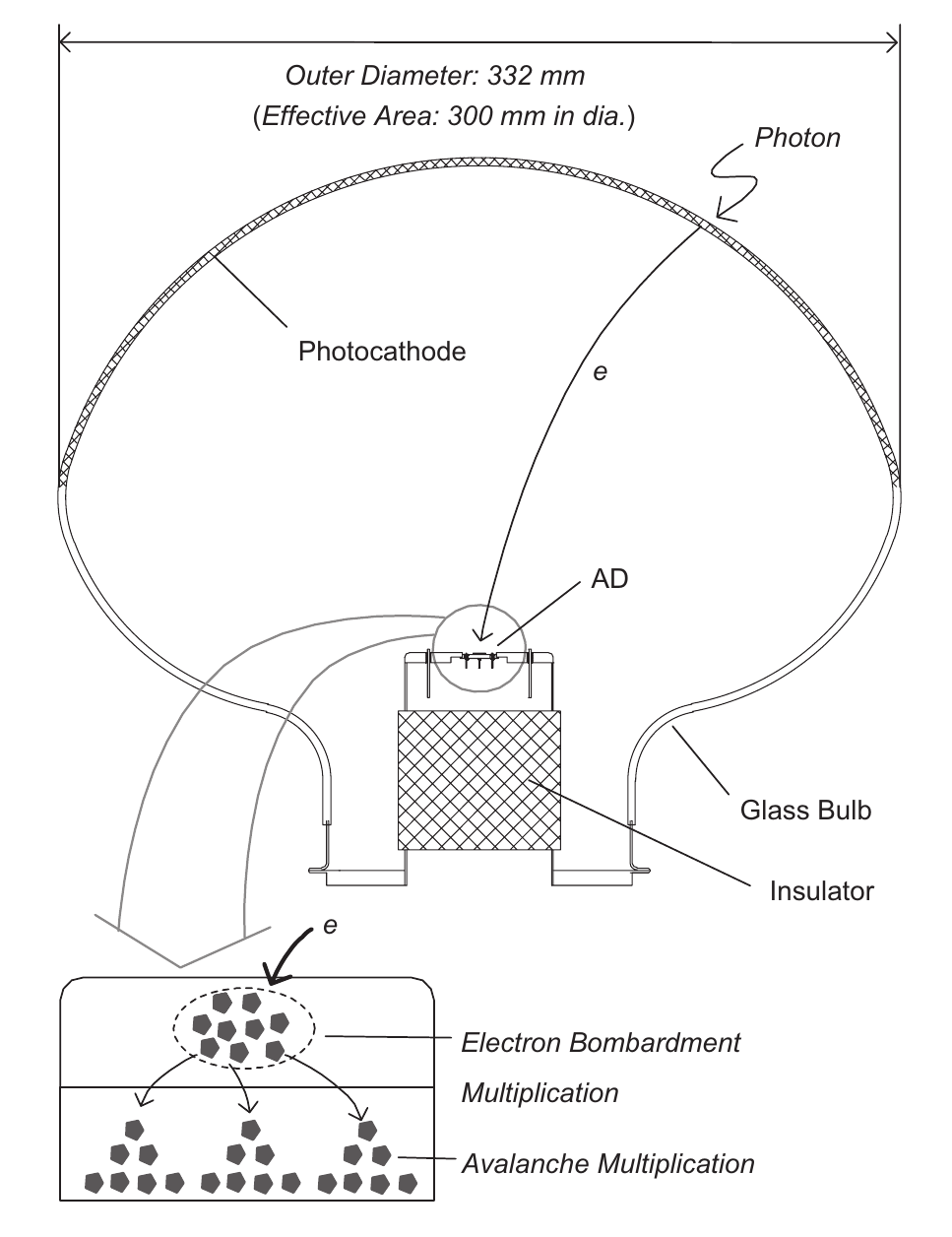}
 \end{center}
 \caption{Outline of the large-aperture Hybrid Avalanche Photo-detector (HAPD) and its principle of operation\cite{hapd}.
 }
\label{hapd}
\end{figure}

A promising large new photo-detector under development\cite{hapd}
 is the Hybrid Avalanche
Photo-detector (HAPD), offering potentially easier construction due to
its simpler structure, and therefore simplified mass production
and lower manufacturing cost.
A 13-inch version has been successfully developed 
and tested\cite{HAPD-2009}.
The internal surface of the HAPD glass envelope is coated with
a photocathode and a light reflector.  The very high voltage ($\sim$ +20 kV)
used for acceleration of electrons toward an avalanche diode results
in a high gain ($\sim 10^5$).  The time resolution is better than
that of a dynode
structure, being about 190 psec, compared to the 1.4 nsec of the
13-inch PMTs.  The elimination of a costly dynode structure,
in favor of an avalanche diode, reduces cost (Figure \ref{hapd}).

\begin{figure}[h]
 \begin{center}
   \includegraphics[width=80mm,keepaspectratio=true]{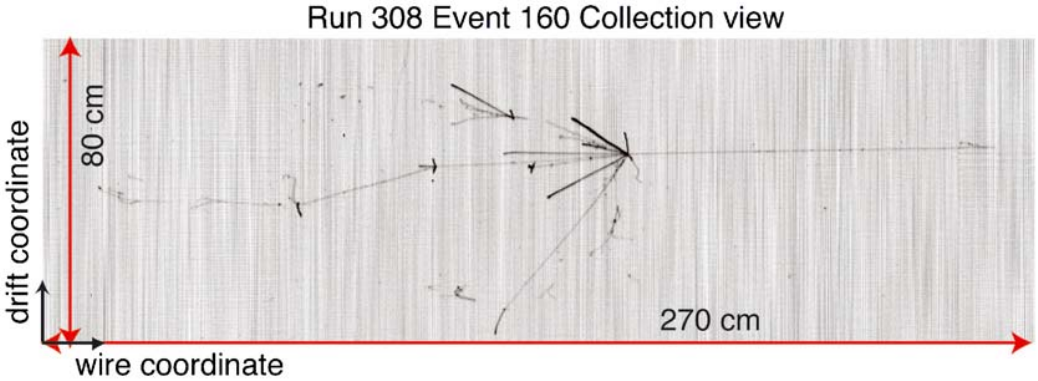}
 \end{center}
 \caption{A hadron shower in the ICARUS detector  \cite{ICARUS04}
 }
\label{icarus}
\end{figure}

A parallel path toward the megaton-class neutrino detector
is offered by the liquid argon TPC.  ICARUS has demonstrated
the potential of this approach, with 300 kilogram chambers (Figure \ref{icarus})\cite{ICARUS04}.
With its low threshold, a 100 kton device is competitive with the
megaton water Cherenkov.  The challenges include
purification, low noise electronics operating cold with signal multiplexing,
design of the vessel including materials and insulation,  underground siting,
and costs\cite{neutrino}.

\subsection{Neutrinoless Double Beta Decay}

Neutrinoless double beta decay holds the potential to establish the
mass scale of the neutrinos. Several 100-200 kg detectors are now being developed.
The major challenge is to minimize backgrounds.
Among the promising efforts are CUORE\cite{cuore}, EXO\cite{exo}, and Majorana\cite{majorana}.
CUORE is a  203 kg $^{130}$Te system composed of
988 TeO$_2$ bolometers. This effort builds on the
11 kg $_{130}$Te CUORICINO effort\cite{cuoricino}.
EXO is developing a 200 kg $^{136}$Xe chamber,
in which ionization and scintillation are measured, and Ba
tagging will be added (EXO-200).
Majorana is working toward a 120 kg $_{76}$Ge experiment.
These, and other efforts, are leading the way toward few milli-eV
$\nu_e$ sensitivity.

\subsection{Dark Matter Direct Detection Techniques}

Given the very small interaction rate expected of dark matter
particles (WIMPs), detector advances strive for
large masses and low thresholds ($\sim$few keV).  The experimental strategy
can employ signals from ionization, scintillation, phonons, or 
a combination.

\begin{figure}[h]
\centering
   \includegraphics[width=80mm]{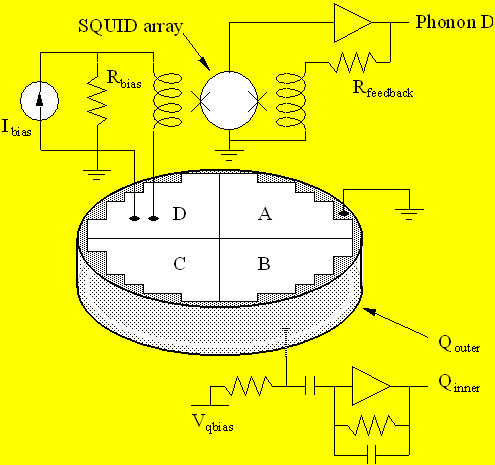}
 \caption{Schematic of the ZIP detector of CDMS, illustrating the
four phonon and two ionization channels along with their readout electronics\cite{cdms}.
 }
 \label{cdms}
\end{figure}

Several bolometers operate at cryogenic temperatures
with state of the art techniques.
CDMS (Cryogenic Dark Matter Search) 
uses germanium and silicon ZIP detectors\cite{cdmszip}  to collect phonons and charge(Figure \ref{cdms}).
With highly segmented readout elements, and good
relative timing of ionization and phonon signals,
good event localization is possible.
A five tower, $\sim$5.5 kg cryogenic assembly has operated in the Soudan mine, and a 25 kg SuperCDMS
is being planned for SNOLAB\cite{cdmsresults}.

\begin{figure}[h]
\centering
   \includegraphics[width=80mm]{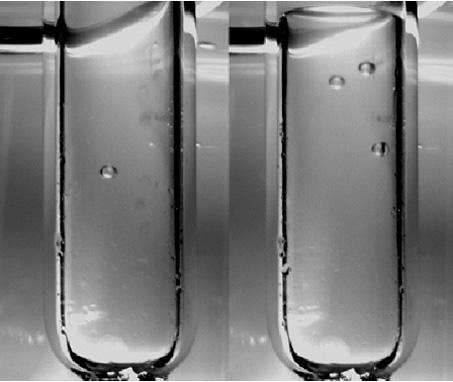}
 \caption{Images from the COUPP bubble chamber.  
A single bubble like that expected from WIMPs (left), and 
several simultaneous disconnected bubbles (right), a signature of neutron recoils. 
\cite{coupp}.
 }
 \label{coupp}
\end{figure}

Examples of other active bolometers include EDELWEISS, CRESST-II, and ROSEBUD.
EDELWEISS has searched for interactions in germanium with Ge/NTD, Ge/NbSi,
and Ge/Interdigit detectors.  EDELWIEISS now operates
thirty kg 
under the Frejus mountain, 
in the underground laboratory of Modane at the French-Italian border.  
CRESST-II
watches 300 grams of CaW0$_4$ crystal in Gran Sasso,
monitoring scintillation and phonons.
ROSEBUD is developing a cryogenic, crystal scintillator of BGO.

The warm liquid bubble chamber \cite{coupp}
 of COUPP (Chicagoland Observatory for Underground
Particle Physics) offers
a different approach in the search for dark matter.  
Two kilograms of CF$_3$I has operated 
successfully (Figure \ref{coupp}).  New 20 and 60 kg chambers are being constructed,
and will go underground in 2010.

\begin{figure}[h]
\centering
\includegraphics[width=80mm]{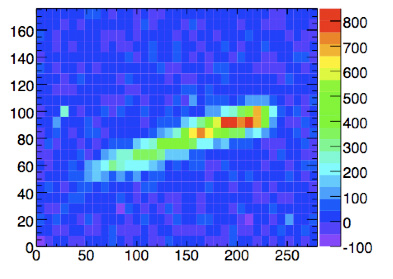}
\includegraphics[width=80mm]{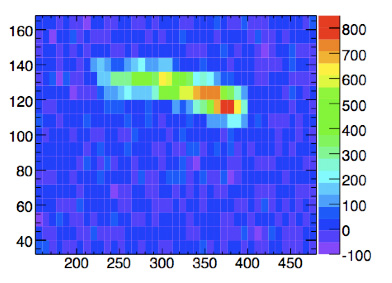} 
\caption{Nuclear recoil candidates of DMTPC\cite{dirdm2} induced by neutrons incident from the
right. 100
pixels correspond to 6 mm.
 }
 \label{DMTPC}
\end{figure}

Another strategy to background reduction is the detector that uses the
recoiling atoms to determine the dark matter particle direction of origin, providing a
suppression of the isotropic background\cite{dirdm}.  Low pressure TPCs are favored
for such ``directional dark matter detectors." CS$_2$ is sensitive to
spin-dependent interactions, and
CF$_4$ and $^3$He  are sensitive to spin-independent interactions.
Different readout approaches are in use.  DRIFT-II uses wire chambers,
with
two 1m$^3$ (CS$_2$) modules now underground.  NEWAGE and MIMAC
are employing  MPGDs, while  DMTPC uses PMT and CCD readout
with a CF$_4$ detector (Figure \ref{DMTPC}).

Nobel liquids (argon, xenon\cite{Aprile:2009dv}, neon) bring many attractive features to the pursuit of dark matter detection:
relatively low cost, ease to obtain, high density as target material,
easily purified due to freeze out of contaminants at cryogenic
temperatures,
very small electron attachment probability,
large electron mobility (that is, large drift velocity for small E-field),
high scintillation efficiency,
and possibility for large, homogenous detectors.
The fundamental limitation and problem is the presence of
radioactive $^{39}$Ar and  $^{85}$Kr.
Efforts include both single phase techniques, and
the two phase approach.
DEAP/CLEAN\cite{Hime:2006zq} and XMASS\cite{Abe:2009zz} are examples of the
single phase work, and XENON\cite{Aprile:2009zzc}, LUX\cite{Fiorucci:2009ak}, and ArDM\cite{Boccone:2008mt}
are examples of the two phase approach.

\subsection{Test Beams}

Test beams are an essential tool for the development of
advanced detectors.  Such beams are important to many
other phases of the particle physics experiment as well, such as prototype
testing, calibrations, etc.  But their value to the early phase of
detector development is essential, and laboratory support for 
this is critical, and appreciated by the community.

\section{Conclusions}
Discoveries in particle physics vitally depend on advances in detector
technology.  The challenges confronting the experimenter today are huge:
greater speed, finer granularity, higher radiation hardness, more exotic materials,
etc.  Many efforts are addressing these challenges.  It is critical to the 
future of the field that these efforts be well funded.  With support, the technology
will advance, providing emerging capabilities critical to future discoveries.

\begin{acknowledgments}
The author is pleased to acknowledge assistance of colleagues
in the preparation of this presentation, including
E. Aprile, M. Breidenbach, A. Bevan, K. Dehmelt, M. Demarteau, B. Fleming, G. Gratta,
D. Hitlin, J. Jaros, H. Ma, G. Rakness, J. Repond, F. Sefkow, A. Seiden, D. Strom, F. Taylor,
J. Timmermans, D. Wark, and A. White.

This work was sponsored in part by the US Department of Energy and the US
National Science Foundation.

\end{acknowledgments}

\bigskip 

\end{document}